\documentclass[12pt,preprint]{aastex}

\usepackage{epsfig}
\usepackage{graphicx, color}

\shorttitle{D. A. Uzdensky}
\shortauthors{Vertical Structure and Coronal Power of MRI Accretion Disks}

\newcommand{\beq}{\begin{equation}}
\newcommand{\eeq}{\end{equation}}

\begin{document}

\title{Vertical Structure and Coronal Power of Accretion Disks Powered by MRI Turbulence.}


\author{Dmitri A. Uzdensky}
\affil{Center for Integrated Plasma Studies, Physics Department, University of Colorado, Boulder, CO 80309 \email{uzdensky@colorado.edu}}

\begin{abstract}
In this paper we consider two outstanding intertwined problems in modern high-energy astrophysics: (1) the vertical thermal structure of an optically thick accretion disk heated by the dissipation of magnetohydrodynamic (MHD) turbulence driven by the magneto-rotational instability (MRI), and (2) determining the fraction of the accretion power released in the corona above the disk. For simplicity, we consider a gas-pressure-dominated disk and assume a constant opacity. 
We argue that the local turbulent dissipation rate due to the disruption of MRI channel flows by secondary parasitic instabilities should be uniform across most of the disk, almost up to the disk photosphere. We then obtain a self-consistent analytical solution for the vertical thermal structure of the disk, governed by the balance between the heating by MRI turbulence and the cooling by radiative diffusion.  
Next, we argue that the coronal power fraction is determined by the competition between the  Parker instability, viewed as a parasitic instability feeding off of MRI channel flows, and other parasitic instabilities. We show that the Parker instability inevitably becomes important near the disk surface, leading to a certain lower limit on the coronal power. 
While most of the analysis in this paper focuses on the case of a disk threaded by an externally imposed vertical magnetic field, we also discuss the zero-net-flux case, in which the magnetic field is produced by the MRI dynamo itself, and show that most of our arguments and conclusions should be valid in this case as well.  
\end{abstract}


\keywords{accretion, accretion disks---magnetic fields---radiative transfer---galaxies: active---X-rays: binaries}


\maketitle


\section{Introduction}
\label{sec-intro}

\subsection{Motivation}
\label{subsec-motivation}

Angular-momentum transport (AMT) in accretion disks is  an important topic in modern high-energy astrophysics, with applications to systems as diverse as Young Stellar Objects (YSOs), accreting stellar-mass compact objects (white dwarfs, neutron stars, and black holes) in galactic binary systems (including X-ray binaries, XRBs), supermassive black holes in active galactic nuclei (AGNs), and even collapsar central engines of long gamma-ray bursts (GRBs). The leading candidate for explaining the observed relatively high AMT levels is the magnetohydrodynamic (MHD) turbulence excited by the magneto-rotational instability (MRI)  \citep{Velikhov-1959, Chandrasekhar-1961, Balbus_Hawley-1991, Balbus_Hawley-1998, Balbus-2003}. Over the past two decades, there have been many analytical and numerical  studies of MRI-turbulent disks [including three-dimensional (3D) MHD simulations, see, e.g., \citep{Brandenburg_etal-1995, Brandenburg_etal-1996, Stone_etal-1996, Armitage-1998, Miller_Stone-2000, Turner-2004, Sano_etal-2004, Hirose_etal-2006, Fromang_Papaloizou-2007, Fromang-2010, Davis_etal-2010, Lesur_Longaretti-2011, Beckwith_etal-2009,  Beckwith_etal-2011,  Simon_etal-2009, Simon_etal-2011, Shi_etal-2010, Guan_Gammie-2011, Bodo_etal-2011, Blaes_etal-2011}]  resulting in a significant progress in our understanding of how these systems work~\citep[see, e.g.,][for review]{Balbus_Hawley-1998, Balbus-2003}.
However, most of these studies have focused, at least until recently, on the basic {\it dynamical} behavior of MRI-driven MHD turbulence in disks, with the primary goal of determining the $r\phi$ component of the turbulent Reynolds and Maxwell stresses that drive accretion. Relatively little attention, in our view, has been paid to the questions of {\it thermodynamics}, in particular, to the vertical thermal structure of MRI-powered disks, governed by the balance between the heating due to turbulent dissipation (which is inextricably linked to the AMT) and the cooling by radiative processes. Correspondingly, most accretion disks simulations have focused on the MHD aspects of the problem, while treating the thermodynamics in a relatively simplified fashion, e.g., considering an isothermal disk, or including optically thin radiative cooling. 
While these approaches have been very helpful in establishing the basic picture of AMT inside disks, they could not address the structure of the upper layers of disks, where the observable radiation is formed, nor the transition to the magnetically-dominated disk corona. One notable exception is the numerical work by~\cite{Hirose_etal-2006} [see also \cite{Turner-2004, Hirose_etal-2009, Shi_etal-2010, Blaes_etal-2011}], who have made the first difficult strides towards a more realistic treatment of MRI-powered radiatively cooled accretion disks. One, of course, has to appreciate the tremendous computational challenge of combining high-resolution 3D MHD simulations, necessary to get the basic magnetohydrodynamics of the problem right, with radiative transfer that spans both optically thick and optically thin regimes.

Another outstanding problem in today's high-energy astrophysics is  understanding the formation of a strongly-magnetized hot tenuous corona above a turbulent accretion disk, in particular, understanding what  governs the fraction of the accretion power that is released in the corona above a relatively cold and dense disk. The observational motivation for our interest in accretion disk coronae (ADCe) stems from the fact that they lie at the base of disk-driven outflows and thus link disks to their winds and jets and also from their role as emitters of high-energy (e.g., X-ray) radiation in many different types of accreting systems. Examples of astrophysical systems were the presence of an ADC has been inferred include: galactic black-hole (BH) XRBs such as Cyg X-1 \citep[e.g.,][]{Bisnovatyi-Kogan_Blinnikov-1976, Liang_Price-1977, Gierlinski_etal-1999, Churazov_etal-2001, DiMatteo_etal-1999};  super-massive BHs in AGNs \citep{Haardt_Maraschi-1991, Haardt_Maraschi-1993, Kawaguchi_etal-2001, Kriss_etal-1999};  and accreting white-dwarf binary systems 
(Cataclysmic Variables, CVs) \cite[e.g.,][]{White_Holt-1982, Church_etal-1998, Naylor_etal-1988, Ramsay_etal-2001}; and YSOs, e.g., T~Tauri stars, \citep{Kwan-1997} (the disk coronal power in these systems is likely to manifest not in X-rays but at lower energies, e.g., UV and optical).

In studying the formation of ADCe, one would like to understand how much magnetic flux and associated magnetic energy emerge from the disk into the overlying corona and what physical processes control the rate and the form of the buoyant magnetic flux emergence. An issue of particular observational importance is the significant variation in the observed relative levels of (coronal) X-ray activity among different spectral states in galactic black-hole X-ray binaries and also among the different types of black-hole accreting systems. In particular, the coronal fraction in AGNs can often be as high as a few tens of percent \citep[e.g.,][]{Svensson_Zdziarski-1994,  Wang_etal-2004}, significantly higher than in galactic black-hole X-ray binaries in the high-soft state. Understanding the reasons for this dichotomy probably requires a detailed quantitative picture of MHD turbulence in a stratified radiation-pressure-dominated disk (which is, after all, the ultimate energy source for coronal activity), with an emphasis on the  production and buoyant rise of magnetic structures~\citep{Blackman_Pessah-2009}. 

A major theoretical development in our understanding of ADC formation was the model of \cite{GRV-1979}, who described the key general magnetic processes leading to the emergence of the corona. Our conceptual theoretical understanding was further advanced by \cite{Tout_Pringle-1992} who outlined a dynamo cycle based on the interplay of the MRI and Parker instabilities.


\subsection{Comments on the Physics of MRI Turbulence}
\label{subsec-comments}

In local shearing-box numerical studies of MRI turbulence, it is customary to distinguish two cases: the case with a finite net vertical magnetic flux, which is conserved in the course of the simulation, and the case with a zero net vertical flux \cite[e.g.,][]{Hawley_etal-1995}.  
Although both cases have been studied extensively in the past, in recent years it seems that the focus has shifted towards the zero net flux case (except for the work by \citealt{Lesur_Longaretti-2011}). We note, however, that the non-zero vertical flux case is still of considerable interest, especially in situations where one is interested in the interaction between the turbulent accretion disk and the large-scale magnetosphere. In particular, one may be interested in the transport (both radial and azimuthal) of the large-scale vertical magnetic flux across the turbulent disk \citep{vanBallegooijen-1989, Lubow_etal-1994, Livio_etal-1999, Spruit_Uzdensky-2005, Rothstein_Lovelace-2008, Beckwith_etal-2009}. One of the most notable examples of such a situation is the magnetic interaction between the disk and the dipole-like magnetic field of the central star, e.g., in the context of Young Stellar Objects and Neutron Star X-ray binaries \cite[e.g.,][]{Uzdensky-2004}. Another example is when one is interested in a large-scale unipolar magnetic field that is believed to be important for launching disk outflows (winds and jets) in YSOs, galactic BH XRBs, AGNs, and collapsar central engines of GRBs. 

In accretion disks threaded by a relatively weak (below equipartition with the gas pressure) vertical magnetic field with a non-zero net flux, gravitational and rotational energy of the accreting matter is first transformed by the MRI into the magnetic and kinetic energy of the so-called channel flows \citep{Hawley_Balbus-1992, Sano_Inutsuka-2001, Sano_etal-2004, Bodo_etal-2008}.
As was shown by \citet[][hereafter GX94]{GX-1994} in the shearing box formulation, these channel flows are exact nonlinear solutions of the MHD equations; they are basically MRI linear eigenmodes that continue to grow exponentially even when they become very large~\citep{GX-1994}. However, as was also shown by~GX94, in reality, channel flows cannot grow indefinitely,  since they themselves become unstable to various secondary {\it parasitic instabilities} --- i.e., instabilities that feed off of the velocity and magnetic field structures produced by the primary MRI mode, which in this context plays the role of a slowly evolving equilibrium. The development of the parasitic modes leads to the disruption of MRI channel flows and a fully-developed turbulence that dissipates energy locally in the disk.
GX94 themselves considered only ideal-MHD parasitic instabilties in the non-stratified case and found two families, one of which is related to the Kelvin-Helmholtz instability.
In principle, however, other parasitic modes are possible, such as non-ideal, visco-resistive modes including the tearing mode \citep{Pessah_Goodman-2009, Latter_etal-2009}. 
It is also important to note that magnetic fields generated by the MRI channel flows are mostly toroidal and are themselves subject to a toroidal-field version of the MRI, which thus should be viewed here as a secondary parasitic instability feeding off the primary MRI mode.

Another limitation of the above-mentioned analytical studies of parasitic modes is that they restricted themselves to considering an unstratified shearing box; this can be thought of as a representation of a small region inside the disk, small compared with the pressure scale-height, $H \sim c_s/\Omega$. 
Thus, these analyses were effectively local not only in cylindrical
radius but also in the height~$z$ above the disk midplane. 
Since the characteristic scale of the fastest-growing MRI mode, 
$l_{\rm mri}$, is proportional to the strength of the vertical 
magnetic field, 
\beq 
l_{\rm mri} \equiv {\lambda_{\rm mri}\over{2\pi}} \sim V_A/\Omega\, ,
\label{eq-l_mri}
\eeq
where $V_A$ is the Alfv\'en velocity corresponding  to the local density and the vertical magnetic field~$B_0$, this approach is justified only if the vertical magnetic field is sufficiently weak, i.e., $V_A \ll c_s$.
The generalization of the channel mode analysis to the case of stratified disks has been developed recently by~\cite{Latter_etal-2010}.

We would like to remark that recently reported numerical evidence from large shearing box simulations suggesting that channel flows are not important in the overall energetics and AMT in accretion disks \citep[e.g.,][]{Longaretti_Lesur-2010} can probably be attributed to a  very restricted definition of channel flows used in these studies, in particular, to restricting it only to axisymmetric modes in the turbulent Fourier spectrum. Of course, if the computational box is sufficiently large in the toroidal direction, then the contribution from axisymmetric (i.e., spanning the entire toroidal extent of the box) modes should indeed be small. This is because any coherent flow and magnetic field  structures, such as channel flows, have only a limited lifetime before they are disrupted by the parasitic modes as discussed above and thus cannot be correlated over distances much larger than about the MRI wavelength. That is, one should not realistically expect channel-mode structures extending (in any direction) to sizes much larger than~$\lambda_{\rm mri}$. Since here we are interested in a situation where $\lambda_{\rm mri} \ll H$, then the contribution of any axisymmetric (or, for that matter, any large-scale, $l \sim H$) structures in the overall energy dissipation and AMT should indeed be inevitably small.  We believe, however, that this does not disqualify any localized channel-mode structures. Locally, on scales of order  $\l_{\rm mri} \ll H$, we can still expect that the picture presented in GX94 basically holds.


\subsection{Objectives of This Paper}
\label{subsec-objectives}


The present paper has two main goals. 
First, we are interested in the vertical structure of a gravitationally stratified accretion disk. For definiteness, we will restrict the present study to the case of a gas-pressure dominated disk, even though it is not directly applicable to inner parts of black-hole accretion disks in the high soft state. Developing a  generalization of our model to the case of a radiation-pressure-dominated disk is left for a future study. 

As mentioned above, most of the previous numerical studies of MRI turbulence (with the exception of~\citealt{Hirose_etal-2006, Hirose_etal-2009, Blaes_etal-2011}) were done using either the isothermal approximation or a simple prescription for optically thin radiative cooling. Standard geometrically thin accretion disks, however, are optically thick, and so one inevitably has to face the optically-thick radiative transfer problem in order to deduce their structure \citep{Shakura_Sunyaev-1973}. 
Thus, the first main goal of the present study is to construct a self-consistent model of the vertical structure of an optically thick accretion disk heated by the local dissipation of MHD turbulence and cooled by radiative diffusion (we shall ignore any external irradiation). 
To attack this problem, for definiteness we will adopt the point of view that the path to turbulence onset lies mostly in the nonlinear disruption of the MRI channel modes by parasitic instabilities, as described above. This assumption will allow us to obtain a concrete vertical profile (namely, flat, see \S~\ref{subsec-Q}) of the magnetic energy dissipation. We note, however, that this  assumption is probably not critical and a similar picture may be developed without an explicit reliance on the concept of parasitic modes. 
In any case, having a specific physically motivated prediction for turbulent dissipation, supplemented with some extra assumptions about the radiative transport properties of the disk (see \S~\ref{sec-disk-structure}), will allow us to construct a full theory of the steady-state%
\footnote{We are here interested in intermediate timescales~$\Delta t$ 
such that $\Omega^{-1}, \Delta t_{\rm rad\ diff} \ll \Delta t \ll \Delta t_{\rm accr}$.}
vertical disk structure, along the lines of the classical analysis of~\cite{Shakura_Sunyaev-1973}.

Our second main objective is to find a way to estimate the fraction of the accretion energy released by the MRI that is not dissipated locally in the disk itself, but instead is transported vertically by buoyantly rising magnetic flux tubes and dissipated in the hot overlying corona. 
{\it In our view, the coronal fraction of the released power is, 
to a large degree, determined by the competion between the  GX94 parasitic 
instabilities and the Parker instability \citep{Parker-1966}, viewed here as another type of parasitic instability} feeding off the horizontal magnetic field  of the primary MRI mode \citep{Foglizzo_Tagger-1995, Tout_Pringle-1992, Blackman_Pessah-2009}. 
As a first step towards investigating this issue quantitatively, we will compare the growth rates of the GX94 and Parker parasitic instabilities and will investigate how their ratio varies with height. 
As we will show in this paper, the Parker instability is slower than other parasitic instabilities in most of the bulk of the disk, but starts to become competitive in the disk's upper layers, leading to a certain lower limit on the coronal power.

In most of the paper, we will focus on the case where the main magnetic field responsible for the MRI is an externally imposed large-scale vertical magnetic field~$B_0$, uniform in~$z$. This is applicable to situations where such a field is relatively strong, namely, stronger than the magnetic field that would be produced in its absence by turbulent dynamo action due to the MRI turbulence itself. We note that, in general, the problem of large-scale dynamo in stratified MRI turbulent disks is still poorly understood and represents a key frontier in the accretion disk research \citep{Brandenburg_etal-1995, Stone_etal-1996, Hawley_etal-1996, Balbus_Hawley-1998, Blackman_Tan-2004, Vishniac-2009, Blackman_Pessah-2009, Simon_etal-2012}.  Nevertheless, as most of the recent zero-net-flux numerical simulations indicate~\citep{Shi_etal-2010, Davis_etal-2010, Guan_Gammie-2011}, the MRI dynamo, by itself, leads to a saturation large-scale magnetic field only at a relatively low level, $B_{\rm dyn}\ll B_{\rm eq}\equiv (8\pi P_0)^{1/2}$, where $P_0$ is the midplane gas pressure.  Moreover, even the small-scale turbulent magnetic field responsible for the AMT in zero-mean flux simulations is also relatively small, of order $B_{\rm turb}^2/8\pi \sim 0.01\, P_0$ \cite[e.g.,][]{Blackman_etal-2008}. Thus, if the external mean vertical field~$B_0$ is larger than this characteristic dynamo field but still weak compared to $B_{\rm eq}$, i.e., if $B_{\rm dyn}^2, B_{\rm turb}^2  \ll B_0^2 \ll B_{\rm eq}^2$, then one may expect  the dynamics to be determined mostly by the super-imposed vertical field~$B_0$. This point of view is also supported by recent numerical studies  by \cite{Lesur_Longaretti-2011} and by \cite{Bodo_etal-2011} and we will adopt it as our starting point in the main part of the paper where we consider the non-zero net flux case. 

We note, however, that the opposite case --- the case where the externally imposed magnetic field is small or absent and where the MRI is driven by the self-generated dynamo magnetic field, --- is also, of course, of considerable interest in astrophysics and we will devote \S~\ref{sec-zero-net-flux} to discussing it.  In particular, we will argue that most of our results for the vertical disk structure obtained in \S~\ref{sec-disk-structure} for the finite-$B_0$ case can also be applied to the zero-net-flux case, with $B_0$ replaced by $B_{\rm dyn}$. Furthermore, we will recover the basic scalings obtained by \cite{Shakura_Sunyaev-1973} for this case and, in addition, will calculate the vertical temperature and density profiles that seem to be in good agreement with the results of numerical simulations by \cite{Hirose_etal-2006}. 
However, we will not attempt to estimate the coronal power fraction for the zero-net-flux case, leaving this task for a future study.

This paper is organized as follows. In \S~\ref{sec-picture} we describe the basic idea and our overall approach to the problem. In \S~\ref{sec-disk-structure} we present the calculation of the internal vertical thermal structure of the disk threaded by a nonzero net vertical magnetic flux.
Next, in \S~\ref{sec-limitations} we consider the upper layers of the disk and investigate how close to the disk surface various assumptions of our model break down.
Then, in \S~\ref{sec-parker} we estimate the growth rate of the Parker instability  and compare it with that of the GX parasitic instabilities, as a function of height; this allows to estimate the coronal fraction of the accretion power as a function of the system's parameters. 
Then, in \S~\ref{sec-Mdot} we discuss the implications of our model for the accretion rate
and for the longer-term evolution of the accretion disk.
The zero-net-flux case is discussed in \S~\ref{sec-zero-net-flux}.  
Finally, we present our conclusions in~\S~\ref{sec-conclusions}.


\section{The Overall Physical Picture}
\label{sec-picture}

The overall physical picture we have in mind can be described as follows.

For MRI to be active in the first place, the vertical magnetic field must be relatively weak, with the corresponding magnetic pressure less than the gas pressure~$P$ in most of the disk,  $B_0^2/8\pi < P$.   
[As mentioned above, for simplicity we ignore radiation pressure in the present work.]
Furthermore, following the previous studies \citep{GX-1994, Pessah_Goodman-2009, Latter_etal-2009}, for simplicity we shall consider the case  when this field pressure is not only weak, but very weak compared with the gas pressure: $B_0^2/8\pi \ll P$. Having the small parameter $\beta^{-1} \equiv B_0^2/8\pi P \ll 1$ at our disposal will allow us to make several important simplifications.

First, in the spirit of GX94 and \cite{Pessah_Goodman-2009}, we assume that the growth of MRI is checked by the development of the parasitic instabilities. 
This assumption allows us to get on a direct path towards evaluating the MRI turbulent energy dissipation rate.%
\footnote{We note, however, that our model should not, in our view, rely critically on our assumptions about the role of parasitic modes in setting the saturation level of MRI turbulence. We believe that essentially similar results can be obtained invoking some other physical picture of the saturation mechanism, e.g., turbulent diffusivity (J.~Goodman, private communication).  }
Once the parasitic instabilities take over, fully-developed MHD turbulence sets in and destroys the horizontal magnetic field of the primary MRI mode. As a result, the magnetic and kinetic energy of the primary MRI mode are dissipated by turbulent cascade on a timescale of order the dynamical time, i.e., $\Omega^{-1}$.
The typical maximum amplitude to which the MRI channel flows are able to grow can be  represented by the characteristic horizontal magnetic field $B_{\rm hor, sat}= b_{\rm sat} B_0$.
It can be estimated from the condition that the growth rate of the fastest-growing parasitic mode is comparable to the growth rate of the fastest growing (and hence the most relevant) primary MRI mode, which is essentially a numerical constant times~$\Omega$, independent of~$\beta$.
The maximum growth rate of the GX94 parasitic  instabilities is naturally
proportional to the amplitude of the primary mode, i.e., $\gamma_{GX}\sim b\Omega$. Therefore, the two growth rates become equal at a certain finite critical value $b=b_{\rm sat}$, generally of order one (even though the GX94 theory was formally developed under the assumption that $b\gg 1$).
Importantly, within the framework of the GX94 model (unstratified disk with~$\beta \gg 1$, ideal-MHD incompressible motions, etc.), there are really no additional parameters on which the dimensionless saturation amplitude $b_{\rm sat}$ could depend. It is then natural to take it to be just a constant number of order unity. In particular, it should be the same for all heights in the disk, as long as we are not close enough to the disk surface, where the assumptions of the model break down (see~\S~\ref{sec-limitations}).

Another reason why the assumption $B_0^2/8\pi \ll P$ is important is that it justifies the locality (in~$z$) of our picture of MRI turbulence. Indeed, the communication speed in the vertical direction is only~$V_A$, and the lifetime of channel modes limited by the disruption by the parasitic instabilities is of order $\Omega^{-1}|\log b_{\rm sat}|$. Any causal connection established during this time extends only over a vertical distance of order~$l_{\rm mri} |\log b_{\rm sat}| \sim l_{\rm mri}$, which is smaller than the gas pressure scale height $H$ by a factor $\beta^{1/2}$. This means that,  as long as our assumption $\beta \gg 1$ holds, channel flows that develop at substantially different heights interact with each other only weakly and hence can be considered separately. In other words, MRI at different heights in the disk develops independently of what happens at other heights. Therefore, since we are interested in the vertical structure of the disk including the vertical profile of the MRI turbulence and its energy dissipation rate, we can just regard properties of MRI channel flows and parasitic instabilities as being local in~$z$. One thus can conclude that, under these circumstances,
the exersize of calculating global MRI eigen-modes spanning the entire thickness of a stratified disk may be of purely academic interest.

The dissipated turbulent energy provides the main heating source of the gas in the disk. Because MRI turbulence is distributed over the disk thickness, the effective heating source is also distributed, $Q=Q(z)$. 
In a thermal steady state, this distributed heating is balanced by the sum of turbulent thermal conduction losses (which, as we shall argue below, effectively just modify $Q$  by a constant factor of order unity) and radiative losses; we assume that the latter take place via optically thick radiative diffusion \cite[e.g.,][]{Shakura_Sunyaev-1973}. 
Solving the corresponding vertical radiative transfer problem yields a full self-consistent
vertical structure of the disk (see~\S~\ref{subsec-structure}).

Having this solution at hand will help us obtain a lower estimate for the coronal power fraction (see~\S~\ref{sec-parker}). 
As mentioned in \S~\ref{subsec-objectives}, our approach to this problem is to regard the Parker instability, which leads to the buoyant rise of the magnetic flux tubes into the corona, 
as a parasitic instability that competes for power with the other (GX94) parasitic instabilities. 
As we shall show in this paper, deep inside the disk, at heights $z \leq H\sim  c_s/\Omega$, 
the Parker instability growth rate is smaller than the characteristic GX94 growth rate, roughly by a factor of $\beta^{1/2}\gg 1$. This means that in the bulk of the disk, the MRI channel flows are destroyed by the GX94 parasitic instabilities well before the Parker instability can develop. Correspondingly, most of the energy density associated with the MRI channel flows ($\sim B_{\rm hor}^2/8\pi \sim b^2 B_0^2$) goes into feeding the resulting local MHD turbulence and is 
dissipated locally in the disk.%
\footnote{We note, however, that a small fraction of this energy may still escape magnetically into the corona if there is an efficient inverse-cascade dynamo producing large enough magnetic structures, e.g., \cite{Blackman_Pessah-2009, Simon_etal-2012}.} 
However, in the upper layers of the disk, as one approaches the disk's photosphere, the Parker instability starts to compete effectively with the other parasitic instabilities and hence one can expect a significant fraction of the MRI energy in this region to be transported up into the corona.

Whether the assumptions on which the above arguments are built are actually valid, will need to be investigated by carefully designed numerical simulations of MRI turbulence in a stratified accretion disk, including full gas thermodynamics with radiative transfer. Since the present paper provides specific predictions for the disk structure and the coronal power, we hope that it will motivate such studies in the future.


\section{Vertical Structure of a Disk Threaded by a Net Vertical Magnetic Field}
\label{sec-disk-structure}



\subsection{Vertical Profile of Energy Dissipation}
\label{subsec-Q}

The first important conclusion we can derive from the arguments in the preceding section is that,  as long as the parasitic instabilities dominate, the volumetric dissipation (i.e., heating) rate should be independent of height~$z$ within the disk. 

Indeed, the vertical magnetic field is independent of~$z$, but the density decreases with height because of stratification. This means that $V_A$ increases with $z$, and so does $l_{\rm mri}$. Then, both the typical vertical and horizontal extent of MRI channel modes, of order $l_{\rm mri}$ and $l_{\rm mri} b_{\rm sat}$, respectively, increase with height; however, as long as they remain $\ll H$, the typical life time of the channel modes, and the local turn-over timescale of the MRI turbulence remain constant, of order $\Omega^{-1}$. The typical horizontal magnetic field component also remains independent of height, of order $B_{\rm hor}\sim B_0\, b_{\rm sat}$. The overall local dissipation rate per unit volume can then be estimated as 
\beq
Q \simeq \gamma_{\rm mri} \, {{B_{\rm sat}^2}\over{8\pi}}\,  
\log^{-1}[B_{\rm sat}/\delta B_{\rm hor}(0)]
\sim \Omega\, {{B_0^2}\over{8\pi}}\, b_{\rm sat}^2 \, . 
\label{eq-Q}
\eeq
Here the logarithmic factor represents the number of $e$-foldings 
needed to grow from some initial perturbation $\delta B_{\rm hor}(0)$ 
to the saturation amplitude~$B_{\rm sat}$. In a realistic situation, 
of course, one does not expect the magnetic field to return exactly 
to a pure vertical-field state; therefore, in the following analysis,
we shall assume that the typical initial perturbation amplitude is of the order 
of~$B_0$ itself, and thus will ignore this logarithmic factor.

Thus we now see that, because the background vertical field threading the disk, $B_z=B_0$, is 
independent of height~$z$ in the case under consideration, the above volumetric MRI dissipation rate is also independent of height, basically unaffected by the pressure stratification! Importantly, it does not have to follow the gas density or the gas pressure.
That is, the volumetric dissipation rate does not follow mass, as it is sometimes 
assumed, but is just constant, uniform in~$z$,
at least across the main part of the disk, where the magnetic field is still dynamically weak. This expectation is in fact consistent with the results of numerical simulations~\citep{Miller_Stone-2000, Hirose_etal-2006} for the zero-net-flux case, which show turbulent magnetic dissipation that is roughly flat over a couple gas scale-heights (or even slightly peaked at about~$z=2H$).

It is important to note that, since the ultimate source of the heating  is the dissipation of the 
accretion energy, the fact that $Q \sim B_0^2$ implies that the accretion torque is also proportional to the square of the mean vertical field. This is in clear contradiction with the results of a number of numerical simulations reporting a linear scaling of the accretion stress with~$B_0$ \citep{Hawley_etal-1995, Pessah_etal-2007} [note, however, that \cite{Sano_etal-2004} report a $B_0^{3/2}$ scaling]. We believe that this discrepancy may be attributed to the fact that most of these numerical studies were not in the asymptotic regime of interest here, i.e. they lacked the required separation of scales between the disk scale height (or the vertical box size~$L$), the prevailing MRI scale $l_{\rm mri}$, and the dissipative scale (or the grid scale). In other words these studies either were not sufficiently resolved, in the sense that their MRI scale corresponding to the net vertical field $B_0$ were not much larger than the resolution scale, or their MRI scale was not much lower than the scale height (for stratified sims) or the box size (for unstratified sims), which is equivalent to saying that $B_0$ was not sufficiently small compared with the pressure-equipartition field (and hence the corresponding MRI wavelength was not much smaller than the~$H$). In fact, a recent very careful numerical study by \cite{Bodo_etal-2011} demonstrates that indeed the accretion torque indeed scales as $B_0^2$ in the asymptotic regime $\delta \ll l_{\rm mri} \ll L$, and hence supports the point of view advanced in the present paper [see also \cite{Longaretti_Lesur-2010}]. 

Another important point here is that the {\it specific} (i.e., per unit mass) AMT due to MRI turbulence and hence the resulting effective accretion  inflow velocity are not uniform in height: at larger heights, the accretion inward drift velocity is higher.  This is similar to the results of numerical simulations of \cite{Beckwith_etal-2009} where this property of MRI turbulence in a stratified disk provided a mechanism for efficient inward transport of the vertical magnetic flux (see also \citealt{Rothstein_Lovelace-2008}).


\subsection{Vertical Structure of an Optically Thick Disk}
\label{subsec-structure}
 
Once the vertical profile of turbulent dissipation is established, we can determine the vertical structure of the disk, similar to the calculation by \cite{Shakura_Sunyaev-1973} for accretion disks and to standard stellar structure calculations. Here we are interested in a steady-state profiles of the gas temperature and density established on timescales longer than the characeteristic orbital time and the cooling time, but shorter than the overall accretion time (characteristic radial transport timescale).

We also neglect external irradiation of the disk, e.g., by the 
radiation coming from the inner part of the disk or the central star. 

It is widely recognized that in discussing MRI in the presence of 
a mean vertical field, the ratio of the magnetic pressure to the gas
pressure, $\beta^{-1} \equiv B_0^2/8\pi P$, is an important parameter. 
However, it is important to realize that the gas pressure profile in 
the disk is not just some arbitrary prescribed function, but needs to 
be determined self-consistently using the physical laws, namely the 
energy transport and the vertical hydrostatic balance. This point was 
made by Hirose~et~al. (2009), and here we adopt this point of view. 
That is, we cannot prescribe 
the temperature and density profiles of the disk; they are to be determined 
as part of the overall problem. Instead, we can prescribe as fixed only 
the those quantities that evolve only relatively slowly, namely, due to 
the radial transport associated with the accretion process itself; these
quantites are then conserved on the timescales of interest to us here.
In the present problem there are basically two such quantities:
the disk surface density of mass, $\Sigma$, and that of the vertical 
magnetic flux, i.e., the vertical magnetic field~$B_0$. In addition,
we can prescribe the local disk rotation rate~$\Omega$, and 
the parameters describing the radiative transfer, e.g., the scattering 
opacity~$\kappa$, which for simplicity we assume to be constant, see below. 
Thus, $\Sigma$, $B_0$, $\Omega$, and $\kappa$ are the only input parameters determining the disk vertical structure in our model. 

It will also be convenient to define a important dimensionless parameter ---  
the disk's optical depth measured from infinity to the disk midplane:
\beq
\tau_{\rm tot} \equiv {1\over 2}\, \Sigma \kappa \, ,
\eeq
We assume the disk to be optically thick, $\tau_{\rm tot} \gg 1$.

Another important dimensionless parameter in our problem is the midplane plasma-$\beta$ parameter, defined with respect to the superimposed vertical field~$B_0$: 
\beq
\beta_0 = \beta(0) \equiv {{8\pi P_0}\over{B_0^2}} \, ,
\eeq
where $P_0 \equiv P(z=0)$. 
This parameter characterizes the dynamical importance of the vertical magnetic field, relative to the gas pressure. In our model we assume that $\beta_0 \gg 1$. Note that since we haven't yet computed the midplane plasma pressure, we cannot, at this stage, express $\beta_0$ in terms of our principal input parameters; this will be done at the end of this section.

Now let us solve for the vertical structure based on the hydrostatic balance and the radiative
energy transport. This calculation is similar to the classical analysis by \cite{Shakura_Sunyaev-1973} and is also similar to the traditional analyses of the radiative stellar structure except that it is performed in plane, rather than spherical, geometry. One important difference between our analysis and that of \cite{Shakura_Sunyaev-1973} is that here we are able, within the framework of our model, to obtain explicit analytical expressions for the dependences of the plasma density~$\rho(z)$ and temperature~$T(z)$ on the height~$z$ within a gas-pressure-dominated disk. 

We shall start with the condition of hydrostatic pressure balance.
Neglecting magnetic and radiation pressure support compared to the thermal gas pressure,  
and assuming that gravity is from the central point mass, we have (for a fully ionized hydrogen plasma):
\beq
{dP\over{dz}} = {2\over{m_p}}\,{{d(\rho k_B T)}\over{dz}} = g_z \rho = -\, \Omega^2 z \rho \, .
\label{eq-pressure-balance-1}
\eeq

That is,
\beq
{4\over{m_p}}\,{{d(\rho k_B T)}\over{d(z^2)}} = -\, \Omega^2  \rho \, .
\label{eq-pressure-balance-2}
\eeq

Next, we need to supplement this equation by the vertical heat balance equation that reads, in a steady state,
\beq
Q = {d\over{dz}} \, (F_{\rm rad} + F_{\rm turb}) \, ,
\eeq
where $F_{\rm rad}$ is the radiative energy flux and $F_{\rm turb}$ is the effective vertical heat flux due to the MRI turbulence itself.  The latter can be estimated roughly as a diffusive flux 
$ F_{\rm turb} (z) =  -\, D_{\rm turb} \, d(3 n k_B T)/dz $, 
with an effective MRI-turbulent diffusion coefficient $D_{\rm turb} \sim \l_{\rm mri} V_A/3 $.  
Then, ignoring factors of order unity, we can write:  
$ F_{\rm turb}(z) \sim -\, \l_{\rm mri} V_A \, d P/dz \sim -\, \Omega^{-1}\, V_A^2\, d P/dz $.
Substituting $dP/dz$ from equation~(\ref{eq-pressure-balance-1}), we get 
\beq
F_{\rm turb}(z) \sim  \Omega\, V_A^2\,  z \rho = \Omega\, {{B_0^2}\over{4\pi}}\,  z \, , 
\eeq
and hence, since both $\Omega$ and $B_0$ are constant in~$z$, 
\beq
{dF_{\rm turb}\over{dz}} \sim \Omega\, {{B_0^2}\over{4\pi}} = {\rm const} \, . 
\eeq
Thus, we see that the effective cooling rate due to the vertical heat transport by the MRI-driven turbulence is basically the same as the MRI turbulent heating rate~$Q$ (see eq.~[\ref{eq-Q}]), apart from a constant numerical coefficient of order unity: $dF_{\rm turb}/dz = f_{\rm turb} Q$; in particular, importantly, it is constant in~$z$. This allows us to combine both of the effects of the MRI turbulence --- the heating by turbulent dissipation and cooling by vertical turbulent transport --- into one single term, the reduced heating rate
\beq
Q' = Q - {dF_{\rm turb}\over{dz}} = \eta Q = {\rm const} \, ,
\eeq
where we took into account our result that $Q'(z)$ is constant (see \S~\ref{subsec-Q}) 
and where we introduced a constant dimensionless factor $\eta \equiv 1-f_{\rm turb} \lesssim 1$ to account for the reduction, due to the turbulent heat transport, of the required radiative losses. 

Correspondingly, the energy balance equation now becomes
\beq
dF_{\rm rad}/dz = Q' = {\rm const} \Rightarrow F_{\rm rad}(z) = Q' \, z \, , 
\eeq

In the radiation diffusion approximation, the vertical radiative energy flux is
\beq
F_{\rm rad} = {c a\over 3}\, {{dT^4}\over{d\tau}} = 
-\, {c a\over 3}\, {{dT^4}\over{d z }}\, \lambda_{\rm ph} = 
-\, {c a\over 3}\, {{dT^4}\over{\rho\kappa d z }} \, ,
\eeq
where $a=4\sigma_{SB}/c = \pi^2 k_B^4/15\hbar^3 c^3 = 
7.566\times 10^{-15}\, {\rm erg\ cm^{-3}\ K^{-4}}$ 
is the radiation constant, $\sigma_{SB}$ is the Stefan-Boltzmann constant,  
and $\lambda_{\rm ph} = -dz/d\tau = (\rho\kappa)^{-1}$ is the photon mean free path.  This allows us to  write our energy balance equation as
\beq
{2 c a\over3}\, {{dT^4}\over{d(z^2)}} = -\, \rho\kappa Q' \, .
\label{eq-rad-transfer}
\eeq

From the two equations~(\ref{eq-rad-transfer}) and~(\ref{eq-pressure-balance-2}) we get:
\beq
{2c a\over 3}\, dT^4 = -\, \kappa Q' \rho d(z^2) = 
{{\kappa Q'}\over{\Omega^2}} \, {4\over{m_p}}\, d(\rho k_B T)\, .
\label{eq-dT4-dP}
\eeq

In general, the opacity should be a function of the local plasma parameters such as the temperature, but for simplicity and definiteness in this study we shall assume that it is constant, 
$\kappa={\rm const}$ (as it would be for the case of electron scattering, for example). 
Investigation of more realistic and complicated opacities, including the transition to the free-free opacity $\sigma_{\rm ff} \sim \rho T^{-7/2}$ in the upper, colder layers of the disk, is left for future work. 

We can now  immediately integrate equation~(\ref{eq-dT4-dP}) to obtain an algebraic relationship between density and temperature:
\beq
\rho(z) = {{c a \Omega^2 m_p}\over{6\kappa Q' k_B}}\, T^3(z) \equiv A T^3\, ,
\label{eq-rho-T}
\eeq
where we defined 
\beq
A \equiv  {{ca\Omega^2 m_p}\over{6\kappa Q' k_B}} = {{2\sigma_{\rm SB} \Omega^2 m_p}\over{3\kappa Q' k_B}}\, ,
\eeq
and where we neglected the integration constant by assuming  that $T(z)$ and $\rho(z)$ effectively become small together at the disk's photosphere. [We assume that the disk is not subject to any substantial external irradiation (such as that coming from the central star for example), which would invalidate the above boundary condition].

At this point we can check that a disk with this structure is always convectively stable for a gas adiabatic index equal to $\gamma_{\rm ad}=5/3$. Indeed, because $\rho(z)\sim T^3(z)$, it follows that 
\beq
\nabla_{\rm rad} = \biggl({{d\log T}\over{d\log P}}\biggr)_{\rm rad}
= {1\over 4} < \nabla_{\rm ad} = {{\gamma_{\rm ad}-1}\over{\gamma_{\rm ad}}} =
{2\over 5} \, .
\eeq
This means that the disk is stable to thermal convection and hence our assumptions regarding the nature of vertical heat transfer are justified.

Substituting equation~(\ref{eq-rho-T}) into~(\ref{eq-pressure-balance-2}), we obtain:
\beq
{{dT^4}\over T^3}  = -\, {{m_p \Omega^2}\over{4 k_B}} \, dz^2 \, , 
\eeq
and hence
\beq
k_B T(z) = k_B T(0) - {{m_p \Omega^2}\over 16}\,  z^2 =
k_B T(0) \, \biggl( 1 - {z^2\over{z_0^2}} \biggr)\, ,
\label{eq-T-z}
\eeq
where $T(0)$ is the midplane temperature and 
\beq
z_0 \equiv  4 c_{s,0}/\Omega
\label{eq-def-z_0}
\eeq
and $c_{s,0}^2\equiv k_B T(0)/m_p$.

Next, using equation~(\ref{eq-rho-T}), we obtain the density profile:
\beq
\rho(z) = \rho(0)\, \biggl( 1 - {z^2\over{z_0^2}} \biggr)^3 = 
A\, T^3(0)\, \biggl( 1 - {z^2\over{z_0^2}} \biggr)^3\, .
\label{eq-rho-z}
\eeq
That is, in this model the disk has a sharp surface 
(similar to stars) at a finite height~$z=z_0$ above midplane. 
In the following, it will sometimes be convenient to normalize $z$ by~$z_0$,
i.e., to use the dimensionless height variable
\beq
\zeta \equiv z/z_0 \, .
\eeq

The midplane values of temperature and density are then determined 
from the condition that the surface density has to be equal to a 
prescribed value:
\beq
2\, \int\limits_0^\infty  \rho(z) dz = \Sigma \, .
\eeq
Thus we get
\beq
\Sigma = 2 z_0 \rho(0)\, \int\limits_0^1 (1-\zeta^2)^3\, d\zeta 
\equiv C_1 z_0\,  \rho(0) \, , 
\eeq
where $C_1 \equiv 2 \int_0^1 (1-\zeta^2)^3 d\zeta = 32/35$. Of course, our model is too crude to consider any factors of order~1 meaningful, but we'll keep them here anyway. 
Thus, the total optical depth from infinity to the midplane of the disk and the central density are related via
\beq
\tau_{\rm tot} = 
{1\over 2}\, \Sigma \kappa = {C_1\over 2}\, z_0 \rho(0) \kappa =
{16\over{35}}\, z_0 \rho(0) \kappa \, .
\eeq

Remembering the definition~(\ref{eq-def-z_0}) of~$z_0$, we then get the central temperature in terms of the primary input parameters: 
\begin{eqnarray}
T_0 \equiv T(0) &=& \biggl[\, {35\over{128}}\, A^{-1} 
\Sigma \Omega\, (m_p/k_B)^{1/2}\biggr]^{2/7}  \nonumber \\
&=&  \biggl[\, {105\over 128}\, {Q'\over{\Omega}}\, 
(k_B/m_p)^{1/2}\, {{\tau_{\rm tot}}\over{\sigma_{\rm SB}}}\, \biggr]^{2/7}  = 
\biggl[\, {105\over 128}\, {{{B_0^2 b_{\rm sat}^2} \eta }\over{8\pi}}\, 
(k_B/m_p)^{1/2}\, {{\tau_{\rm tot}}\over{\sigma_{\rm SB}}}\, \biggr]^{2/7} \, ,
\label{eq-T0}
\end{eqnarray}
where we used $Q' = \eta\, \Omega\, B_0^2 b_{\rm sat}^2/{8\pi}$ as the fiducial reduced characteristic MRI heating rate, to get the last expression. It is interesting to note that the above expression for the midplane disk temperature involves only the vertical magnetic field pressure and the total disk optical depth, plus some universal physical constants. In particular, one can see that $T(0)$ does not depend on the orbital frequency~$\Omega$. This can be understood qualitatively as follows. The total dissipation rate per unit disk area is of order $z_0 Q' \sim (c_{s,0}/\Omega) \, (\eta \Omega \, B_0^2 b_{\rm sat}^2/8\pi) \sim T^{1/2}(0) \, \eta \, (B_0^2 b_{\rm sat}^2/8\pi)$ --- independent of~$\Omega$. 
On the other hand in this model this dissipated power is emitted from the disk's photosphere as thermal emission: 
$F \sim \sigma_{\rm SB} T_{\rm ph}^4  \sim \sigma_{\rm SB} T^4(0) \tau_{\rm tot}^{-1}$. 
Comparing these two expressions, we immediately find the above scaling of $T(0)$ with $B_0^2$ and $\tau_{\rm tot}$. 

Numerical estimate yields: 
\beq
T(0) \simeq 220 \, {\rm K} \ \tau_{\rm tot}^{2/7}\, [\eta\, B_0^2 b_{\rm sat}^2/{8\pi}]^{2/7} \, .
\eeq
Thus, in the case of galactic BH XRBs, for example, with a typical $B_0\sim 10^7$~G, 
we get $T(0) \simeq 10^6\, {\rm K}\ \tau_{\rm tot}^{2/7}$, which is not unreasonable.
However, in reality we, of course, do not expect our present model to apply in the inner parts of a black hole accretion disk because of the predominance of radiation pressure there.

With the above result for the midplane disk temperature at hand, we can estimate other important disk quantities, e.g., the speed of sound 
\beq
c_{s,0} = \sqrt{{k_B T(0)}\over{m_p}} \sim \biggl[{105\over{128}}\, {Q'\over{\Omega}}\, 
\biggl({k_B\over{m_p}}\biggl)^4\,  {{\tau_{\rm tot}}\over{\sigma_{\rm SB}}}\, \biggr]^{1/7} \, ,
\label{eq-c_s0}
\eeq
and the disk half-thickness:
\beq
z_0 = 4\, c_{s,0}/\Omega = 2\, \Omega^{-8/7} \, 
\biggl[105\, \biggl({k_B\over{m_p}}\biggl)^4\, {{Q' \tau_{\rm tot}}\over{\sigma_{\rm SB}}}\, \biggr]^{1/7} \, .
\eeq

Next, the central density becomes
\beq
\rho_0 \equiv \rho(0) = A T_0^3  =  
{{(35\, \tau_{\rm tot})^{6/7}}\over{32\, \kappa}} \, 
\biggl[ {{\sigma_{\rm SB}\, \Omega^8}\over{3\, Q'}}\, \biggl({m_p\over{k_B}}\biggl)^4\, \biggr]^{1/7} \, .
\label{eq-rho0}
\eeq

The gas pressure profile is given by 
\beq
P(z) = {2\rho k_B T\over{m_p}} = {2Ak_B \over{m_p}}\, T^4(z) = 
{2A k_B\over{m_p}}\, T_0^4 \, \biggl[1-{{z^2}\over{z_0^2}}\biggr]^4 
= P_0\, \biggl[1-{{z^2}\over{z_0^2}}\biggr]^4 \, ,
\label{eq-P-z}
\eeq
where the central gas pressure is
\begin{eqnarray}
P_0 \equiv P(0) &=& 2\, \rho_0 k_B T_0/m_p = 
{35\over{128}}\, \Omega \, \Sigma \, \sqrt{k_B\over{m_p}} \, 
\biggl[105\, {Q'\over\Omega} \, \sqrt{k_B\over{m_p}}\, {\tau_{\rm tot}\over{\sigma_{\rm SB}}} \, \biggr]^{1/7} \nonumber \\
&=& {35\over{128}}\, \Omega \, \Sigma \, \sqrt{k_B\over{m_p}} \, 
\biggl[ 105\, {{\eta\, B_0^2 b_{\rm sat}^2}\over{8\pi}} \, \sqrt{k_B\over{m_p}}\, {\tau_{\rm tot}\over{\sigma_{\rm SB}}} \, \biggr]^{1/7}  \, .
\end{eqnarray}
It is interesting to note a very weak dependence of $P_0$ on $Q'$ and $B_0$, as well as 
on the opacity~$\kappa$, and a relatively strong dependence on $\Sigma$ and~$\Omega$.

Likewise, the local plasma-$\beta$ parameter associated 
with the vertical field $B_0$ is
\beq
\beta(z) \equiv 8\pi P/B_0^2 = \beta_0\, \biggl[1-{{z^2}\over{z_0^2}}\biggr]^4 \, , 
\label{eq-beta-z}
\eeq 
with $\beta_0 \equiv\beta(0)\gg 1$ 
\beq
\beta_0 \equiv \beta(0)= {P_0\over{B_0^2/8\pi}} \sim
{35\over{128}}\, \Omega \, \Sigma \, \biggl(\eta\, {{B_0^2}\over{8\pi}}\biggr)^{-6/7} \, \sqrt{k_B\over{m_p}} \, 
\biggl[ 105\, b_{\rm sat}^2\, \sqrt{k_B\over{m_p}}\, {\tau_{\rm tot}\over{\sigma_{\rm SB}}} \, \biggr]^{1/7} 
\label{eq-beta_0}
\eeq
being the midplane value; our model assumes $\beta_0 \gg 1$.

This completes the calculation of the interior vertical structure of an optically thick MRI-heated gas-pressure-dominated accretion disk threaded by a relatively weak vertical magnetic field.


\section{Breakdown of the Model near the Disk Photosphere}
\label{sec-limitations}

The above model of the disk interior vertical structure relies on several assumptions that should be well justified deep within the disk but are expected to break down close to the disk surface, 
as $z\rightarrow z_0$. This situation is similar to the one in stellar structure calculations as one approaches the star photosphere. 
These assumptions can be cast in terms of a certain ordering of the relevant physical length scales which can be summarized as follows: 
\beq
l_{\rm mri}(z), \lambda_{\rm ph}(z) \ll  H(z) < \Delta z < z_0\, .
\eeq
Here, $H(z) \equiv -(d\ln P/dz)^{-1}$ is the disk local gas pressure scale height, 
and $\Delta z \equiv z_0-z$ is the geometrical depth, i.e., the distance from the disk edge. 

In this section we shall estimate how rapidly each of these scales varies as one approaches the disk photosphere and will thus estimate at what optical depth the individual components of the above ordering break down.

Here we are interested in the region near the disk's surface: 
\beq
\Delta z \equiv z_0-z \ll z_0 \, , 
\eeq
or, equivalently, 
\beq
\Delta \zeta \equiv \Delta z/z_0 \ll 1 \, .
\eeq

In what follows, it will often be convenient to use the optical depth coordinate 
$\tau(z) = \int_z^{z_0} \kappa \rho(z)dz$, along with $z$ itself; the two quantities are related to each other by equation~(\ref{eq-rho-z}). 
In particular, near the disk edge, $\Delta \zeta \ll 1$,  we have 
\beq
\tau(z\simeq z_0) \simeq  
2 \kappa \rho(0) z_0 \, \Delta \zeta^4 = {35\over{8}}\, \tau_{\rm tot}\, \Delta \zeta^4 \, .
\label{eq-tau-Delta_z}
\eeq
Since equation~(\ref{eq-rho-z}) was derived for the optically thick region, the above expression is valid only for $\tau \gg 1$, i.e., for $1 \gg \Delta \zeta \gg 0.7 \tau_{\rm tot}^{-1/4}$. 

It is also interesting to note that because the gravity is nearly constant near the disk edge, $g_z(z\simeq z_0) = -\, \Omega^2 z \simeq -\, \Omega^2 z_0 = {\rm const}$, the vertical 
hydrostatic pressure balance, equation, eq.~(\ref{eq-pressure-balance-1}), yields
\beq
P(\tau)\simeq \tau |g_z(z_0)|/\kappa \simeq \tau \Omega^2 z_0/\kappa \, .
\eeq
That is, the variation of the optical depth $\tau$ with the geometrical depth $\Delta z$ is the same as that of the gas pressure in this region, and hence $\tau$ can be used as a proxy for the pressure {({\it c.f.} \cite[e.g.,][]{Shakura_Sunyaev-1973, Hirose_etal-2006}.


Now we are in a position to check the validity of some of 
the assumptions underlying our model. 

{\bf {(i)} Gas scale-height, $H$.}
First, for our model to be valid near the disk surface, we must require
that $\Delta z = z_0-z > H(z)$. 

The gas pressure scale-height, using equation~(\ref{eq-P-z}), is 
\beq
H(z)=  -\, \biggl({dP\over{Pdz}}\biggr)^{-1} = {{z_0^2 - z^2}\over{8 \, z}} =
{z_0\over 8}\, {{1-\zeta^2}\over{\zeta}} \, .
\label{eq-H-general}
\eeq
Near disk edge, $\Delta \zeta \ll 1$, this becomes
\beq
H(z)  \simeq {1\over 4}\, \Delta z \, ,
\label{eq-H-Delta-z}
\eeq
so that the assumption $ H(z) < \Delta z$ is marginally satisfied.


{\bf {(ii)} Photon mean-free path, $\lambda_{\rm ph}$.}
Similarly, because of the radiative diffusion approximation, 
our model is valid only as long as the photon mean free path 
is sufficiently short, i.e.,  $\lambda_{\rm ph} \ll H, \, \Delta z$. 
Using the above expressions, we have
\beq
\lambda_{\rm ph} = {1\over{\rho\kappa}} = 
{1\over{\rho(0)\kappa}}\, (1-\zeta^2)^{-3} \simeq 
{\lambda_{\rm ph,0}\over{8}}\, \Delta\zeta^{-3} \, ,
\eeq
where $\lambda_{\rm ph,0}$ is the photon mean-free path
at the disk midplane:
\beq
\lambda_{\rm ph,0} = {1\over{\rho(0)\kappa}} =  
{16\over{35}} \, z_0 \, \tau_{\rm tot}^{-1} \, .
\eeq
From this we see that near the disk edge
\beq
{\lambda_{\rm ph}\over{H}} \simeq {4\lambda_{\rm ph}\over{\Delta z}} = 
{\lambda_{\rm ph,0}\over{2 z_0}} \, \Delta \zeta^{-4} = 
{8\over{35}}\, \tau_{\rm tot}^{-1}\, \Delta \zeta^{-4} = {1\over{\tau(z)}} \, . 
\eeq
Thus we see that that the condition that $\lambda_{\rm ph}(z)  \ll H(z) < \Delta z$ 
is automatically satisfied as long as we are in the optically thick part of the disk, 
$\tau(z) \gg 1$.


{\bf {(iii)} MRI scale, $l_{\rm mri}$.}
Next, we want to check that $l_{\rm mri}$ is less than $H$ and~$\Delta z$. 
We have:
\beq
l_{\rm mri} = {{V_A}\over\Omega} = 
l_{\rm mri,0}\, \biggl({\rho\over\rho_0}\biggr)^{-1/2} = 
l_{\rm mri,0}\, \biggl(1-\zeta^2\biggr)^{-3/2} \, , 
\eeq
where 
\beq
l_{\rm mri,0} = {z_0\over{2\sqrt{\beta_0}}} \, .
\eeq
is the MRI scale at $z=0$. 

Near the disk edge we then have 
\beq
l_{\rm mri} \simeq  {{l_{\rm mri,0}}\over{2\sqrt{2}}}\, \Delta\zeta^{-3/2} = 
{z_0\over{4\sqrt{2\beta_0}}} \, \Delta\zeta^{-3/2} \, , 
\eeq
or
\beq
l_{\rm mri} \simeq 
{{l_{\rm mri,0}}\over{2\sqrt{2}}}\, \biggl({8\over{35}}\,\bar{\tau}\biggr)^{-3/8} \simeq 
0.31\,  z_0 \, \beta_0^{-1/2}\, \bar{\tau}^{-3/8} \, , 
\label{eq-l_mri-tau}
\eeq
where $\bar{\tau}(z)  \equiv \tau(z)/\tau_{\rm tot}$.  

Thus, we have
\beq
{l_{\rm mri}\over{H(z)}} \simeq {4 l_{\rm mri}\over{\Delta z}} \simeq 
(2\beta_0)^{-1/2} \, \Delta\zeta^{-5/2} \simeq 1.8\, \beta_0^{-1/2} \, \bar{\tau}^{-5/8}\, , 
\label{eq-l_mri-H}
\eeq
and so the condition $l_{\rm mri}(z) \ll H(z)$ is satisfied as long as 
$\Delta \zeta \gg (2\beta_0)^{-1/5}$ or, equivalently, $\bar{\tau} \gg 2.5\, \beta_0^{-4/5}$.


{\bf {(iv)}} 
Finally, the calculation presented in the preceding section neglects the magnetic contribution in the vertical pressure balance, which is valid only as long as the local plasma-$\beta$ is greater than~1. According to equation~(\ref{eq-beta-z}), in the outer layer of the disk, the plasma-$\beta$ parameter can be written as
\beq
\beta(z \rightarrow z_0)  = \beta_0\, (1-\zeta^2)^4 \simeq 16\beta_0\, \Delta\zeta^4 \simeq 
{128\over{35}}\, \beta_0\, \bar{\tau} \, .
\label{eq-beta-tau}
\eeq
Thus, the assumption $\beta\gg 1$ is valid only for $\bar{\tau} \gg (35/128)\, \beta_0^{-1}$. 
At optical depths below this, our model for the disk structure becomes invalid. Notice, however, that for $\beta_0 \gg 1$, the critical optical depth corresponding to $\beta=1$, $\bar{\tau} = (128/35)\, \beta_0^{-1}$, 
 is smaller than the critical optical depth $\bar{\tau} = 2.5\, \beta_0^{-4/5}$ at which $l_{\rm mri}$ becomes equal to the gas scale height~$H$. This implies the condition $l_{\rm mri} < H$ is more restrictive, and hence more important, than the condition $\beta\gg 1$.


\section{Parker Instability and Coronal Power}
\label{sec-parker}

Let us now consider the Parker instability in the disk, 
with an ultimate goal of estimating the coronal fraction
of the accretion power. We wish to remind the reader that
the spirit of our approach is to consider the Parker instability
as a secondary parasitic instability feeding on the horizontal 
magnetic field of the  primary MRI mode \citep[e.g.,][]{Tout_Pringle-1992}, 
and competing with the usual GX94 parasitic instabilities. 
Our approach is to compare the linear growth-rates of these two types of parasitic modes and see under what conditions the Parker instability becomes important.

The linear growth rate of a maximally-unstable parasitic instability scales as $\gamma_{GX}\sim \Omega b$, where $b=B_{\rm hor}/B_0$ is the MRI channel flow's horizontal magnetic field normalized by the vertical magnetic field~\citep{GX-1994}. 
The growth rate of the fastest growing Parker instability can be estimated as 
\beq
\gamma_P(z)\sim V_{A, \rm hor} /H(z) = b V_A/H = \Omega b\, l_{\rm mri}/H \,.
\eeq
Thus, the ratio of the growth rates of the two types of parasitic instabilties is just proportional to the ratio of the MRI scale $l_{\rm mri}$ to the disk local pressure scale height~$H(z)$: 
\beq
{{\gamma_P}\over{\gamma_{GX}}} \sim {{l_{\rm mri}}\over{H}}  \, .
\label{eq-Parker_vs_GX-1}
\eeq

As we see, this ratio is small in most of the disk's volume and hence the GX94 parasitic instabilities win over Parker. However, as one approaches the disk surface the above ratio increases because $H(z)$ starts to decrease due to decreasing temperature and 
increasing gravity, while, simultaneously, the MRI scale $l_{\rm mri}\sim B_0/\Omega\sqrt{4\pi \rho}$ increases due decreasing density. In particular, using our estimate~(\ref{eq-l_mri-H}) near the disk surface, we get
\beq
{{\gamma_P}\over{\gamma_{GX}}} \sim {{l_{\rm mri}}\over{H}} \sim 
\beta_0^{-1/2}\, \bar{\tau}^{-5/8} \, .
\label{eq-Parker_vs_GX-2}
\eeq

Thus, at a characteristic optical depth of order
\beq
\tau_P = \tau_{\rm tot}\, \beta_0^{-4/5} \, 
\label{eq-tau_P}
\eeq
the local MRI scale $l_{\rm mri}$ becomes comparable with~$H$
and hence the the growth rate of the secondary Parker instability 
becomes comparable with that of the GX94 parasitic instabilities.
We conjecture that, in the region above the critical depth $z_P\equiv z(\tau_P)$, the Parker instability starts to compete with other parasitic modes as the main mechanism of 
destroying MRI channel flows.
At the same time, however, we acknowledge that, since the condition
$\lambda_{\rm mri} \ll H$ is no longer satisfied in this region, the classical incompressible 
analysis of GX94 is no longer valid, and the effects of stratification \citep[e.g.,][]{Latter_etal-2010} need to be 
taken into account. Nevertheless, it is important to note that since, according to~(\ref{eq-beta-tau}), the plasma-$\beta$ parameter at $\tau=\tau_P$ is still greater than~1 (it is of order $\beta_P \sim \beta_0^{1/5} \gg 1$), there is still a sizable region above~$z_P$ where MRI, although modified by stratification, continues to operate and leads to release of the gravitational energy. It then seems reasonable to suggest that a substantial  fraction of the accretion energy released at heights above~$z_P$ is transported buoyantly into the corona. Thus, a reasonable estimate for the thickness $\Delta z_P$ of this buoyantly active region is given by the corresponding local MRI length-scale (which is of the order of the local pressure scale height at $\tau=\tau_P$), which can be estimated using eqs.~(\ref{eq-l_mri-tau}) and~(\ref{eq-tau_P}):
\beq
\Delta z_P \sim l_{\rm mri}(\tau_P) \sim z_0\, \beta_0^{-1/5} \ll z_0 \, .
\label{eq-Delta-z_P}
\eeq

Since the reduced volumetric MRI energy-release rate $Q'$ is roughly uniform across the disk (see \S~\ref{subsec-Q}), we come to the conclusion that the fraction~$f$ of the accretion power released in the corona should scale simply as
\beq
f \sim  \Delta z_P /z_0 \sim \beta_0^{-1/5} \, .
\label{eq-f-beta_0}
\eeq

Using equation~(\ref{eq-beta_0}), the dependence of the coronal fraction on the primary input parameters of our model can be expressed as
\beq
f \propto \Omega^{-1/5}\, \Sigma^{-1/5} \, \biggl({{B_0^2}\over{8\pi}}\biggr)^{6/35} \, 
\tau_{\rm tot} ^{1/35} \, .
\eeq

Thus, within the limitations of our theory, the coronal power fraction is rather insensitive to most of the input parameters.  

Recall now that one of the assumptions of our model was that the vertical net field~$B_0$, 
while small, is still larger than the field that would be generated by the large-scale MHD dynamo in the zero-net flux case. This limits the midplane plasma-$\beta$ parameter to something of order~100, with the corresponding coronal fraction no less than $f_{\rm min} \sim \beta_{0,\rm max}^{-1/5} \sim 0.4$.


\section{Mass Accretion Rate}
\label{sec-Mdot}

According to \cite{Shakura_Sunyaev-1973}, the mass accretion rate $\dot{M}$ is related to the 
total energy dissipation in a Keplerian disk via
\beq
2 \int\limits_0^{z_0} Q' dz = 2\, Q' z_0 = {3\over{8\pi}}\, \dot{M} \Omega^2 \, .
\eeq
This allows us to express the mass accretion rate in terms of 
the local quantities in our model:
\beq
\dot{M} = {{64\pi}\over 3}\,
{{B_0^2 b_{\rm sat}^2}\over{8\pi\Omega^2}}\, c_{s,0} \, .
\eeq

Substituting our expression~(\ref{eq-c_s0}) for~$c_{s,0}$ we get 
a relationship between $\dot{M}$ and $B_0$, $\Omega$, and~$\Sigma$ (or $\tau_{\rm tot}$):
\beq 
\dot{M}[B_0(R), \Omega(R), \Sigma(R)] \simeq 
65\, \biggl(\eta\, {{B_0^2 b_{\rm sat}^2}\over{8\pi}}\biggr)^{8/7}\,  \Omega^{-2}
\biggl({k_B\over{m_p}}\biggr)^{4/7}\, 
\biggl({\tau_{\rm tot}\over{\sigma_{\rm SB}}}\biggr)^{1/7} \, . 
\label{eq-Mdot}
\eeq
As we see, the mass accretion rate has a relatively strong dependence on the magnetic field and on the rotation rate, but a rather weak dependence on the total disk column density and optical depth.  This effective insensitivity of $\dot{M}$ to $\tau_{\rm tot}$ can be exploited. 
For example, ignoring the $\tau_{\rm tot}^{1/7}$ dependence and taking the $\Omega(R)$ profile to be Keplerian, $\Omega_K\sim R^{-3/2}$, we get a scaling 
$\dot{M} \sim B_0^{16/7}(R)\, R^3$. Thus, we see that a stationary accretion regime, 
$\dot{M}(R) = {\rm const}$ requires a special particular magnetic field profile, 
$B_0^{\rm stat}(R) \sim R^{-21/16}$, which is in agreement with~\cite{Shakura_Sunyaev-1973}.

In general, equation~(\ref{eq-Mdot}) gives us an equation that governs a longer-term evolution for the gas surface mass density: $\dot{\Sigma}(R,t) = d\dot{M}/2\pi RdR$, relating the evolution of the disk density to the radial distribution of the vertical magnetic flux through the disk. Determining the latter, however, is itself an outstanding problem in modern accretion disk research and is still far from being solved  \citep[e.g.][]{vanBallegooijen-1989, Lubow_etal-1994, Spruit_Uzdensky-2005, Uzdensky_Goodman-2008, Rothstein_Lovelace-2008, Lovelace_etal-2009, Beckwith_etal-2009, Fromang_Stone-2009}.


\section{Vertical Structure of the Disk with Zero Net Flux}
\label{sec-zero-net-flux}

The picture presented in the previous sections was developed under the assumption that the super-imposed net vertical magnetic field, $B_0$, while weak compared with the gas pressure inside the disk, nevertheless dominates over the field $B_{\rm dyn}$ that would be generated by the MRI-driven turbulent dynamo in the absence of~$B_0$. Since, as numerical simulations demonstrate, $B_{\rm dyn}$ is typically indeed relatively weak, corresponding to a central plasma-$\beta$ of order $\beta_{\rm dyn} \sim 100$ \cite[e.g.,][]{Hirose_etal-2006, Davis_etal-2010, Shi_etal-2010}, there is indeed a sizable range of parameters where the condition $B_{\rm dyn}^2 \ll B_0^2 \ll B_{\rm eq}^2$ is satisfied and hence where the above picture applies. 

However, it is also interesting to consider the zero net flux case where there is no externally imposed vertical magnetic field (or where this field is weak compared with~$B_{\rm dyn}$). This case  has attracted a lot of attention in the MRI literature, especially in recent years~\cite[e.g.,][]{Hirose_etal-2006, Fromang_Papaloizou-2007, Pessah_etal-2007, Davis_etal-2010, Shi_etal-2010, Guan_Gammie-2011, Simon_etal-2011}. In this case the magnetic field responsible for driving the MRI is the self-generated turbulent field produced by an MHD dynamo associated with the turbulence itself. One expects both small-scale and large-scale dynamo generating the field on a broad range of spatial scales. The problem of computing analytically the overall spectrum of the resulting magnetic field and its vertical distribution in a stratified disk is an outstanding problem in accretion disk theory. This formidable challenge has so far eluded a full solution, although several important theoretical advances have already been made \cite[e.g.,][]{Lesur_Ogilvie-2008,  Vishniac-2009, Blackman_Pessah-2009, Blackman-2012}.
It is, however, reasonable to expect that the characteristic length scale of the average magnetic pressure (averaged over turbulent fluctuations) is much larger than the dominant MRI wavelength $l_{\rm mri}$ and, especially if a strong large-scale MRI dynamo is active, may even be larger than the gas pressure scale height~$H$. In that case, in line with the arguments presented in~\S~\ref{subsec-Q}, we may expect the volumetric dissipation rate $Q(z) \sim \Omega B^2/8\pi$ to be also roughly uniform in~$z$ within the disk. We note that this conclusion differs drastically from the assumption that $Q(z)$ traces the mass density profile, $Q\sim \rho(z)$, adopted by~\cite{Shakura_Sunyaev-1973}. However, our conclusion is in a good agreement with the results of numerical simulations by \cite{Hirose_etal-2006} and~\cite{Shi_etal-2010}--- the only known to us numerical studies of MRI turbulence in a gas-pressure-dominated disk with optically-thick radiative cooling  --- who report a flat top-hat energy dissipation profile over 3 scale heights on each side of the disk. [The flat magnetic energy and dissipation profiles are also consistent with the findings by \cite{Simon_etal-2011} for isothermal stratified disks.] If this is indeed the case, then we can apply the rest of our analysis presented in \S~\ref{subsec-structure} to the zero-net flux case, except that $B_0^2/8\pi$ should be replaced with $B_{\rm dyn}^2/8\pi \sim \beta_{\rm dyn}^{-1} P_0$. In particular, we still recover the $z$-profiles of $\rho$ and $T$ as those given by equations~(\ref{eq-T-z}) and~(\ref{eq-rho-z}), e.g., a parabolic profile for the temperature $T(z) = T_0\, (1-z^2/z_0^2)$. We find that these profiles agree well with the actual average profiles measured in numerical simulations by \cite{Hirose_etal-2006}  (see their Figs.~2 and~3).  

It is important to note that if the zero-net-flux case the characteristic magnetic field strength can no longer be viewed as an external input parameter but rather has to be determined self-consistently along with the other disk quantities. The only input parameters in this case are $\Omega$ and~$\Sigma$, plus the opacity~$\kappa$ (which we assume to be constant, as is the case of the dominant electron scattering). In lieu of using the magnetic field as an input parameter, we can just use the result $\beta_0 = \beta_{\rm dyn} \sim 100$ obtained from numerical simulations. Substituting this on the left-hand side of equation~(\ref{eq-beta_0}) for $\beta_0$, we can rearrange it to express the characteristic magnetic energy density in terms of $\Omega$, $\Sigma$, and $\kappa$, and then substitute the resulting expression for the magnetic energy into our expressions (\ref{eq-T0})-(\ref{eq-rho0}) for~$T_0$, $z_0$, $\rho_0$, etc. Ignoring factors of order unity we get: 
\beq
{{B^2}\over{8\pi}} \sim \biggl({k_B\over{m_p}}\biggl)^{2/3} \, 
\biggl({\tau_{\rm tot}\over{\sigma_{SB}}}\biggr)^{1/6} \, 
\beta_{\rm dyn}^{-7/6} \, \Omega^{7/6}\, \Sigma^{7/6} \, , 
\eeq
and then 
\beq
T_0 \sim \biggl( {k_B\over{m_p}}\, 
{\tau_{\rm tot}\over{\sigma_{SB}}} \, 
\beta_{\rm dyn}^{-1} \, \Omega\, \Sigma \biggr)^{1/3} \, , 
\eeq
\beq
c_{s,0} \sim \sqrt{T_0\over{m_p}} \sim \biggl( {k_B\over{m_p^4}}\, 
{\tau_{\rm tot}\over{\sigma_{SB}}} \, 
\beta_{\rm dyn}^{-1} \, \Omega\, \Sigma \biggr)^{1/6} \, , 
\eeq
\beq
z_{0} \sim c_{s,0}/\Omega    \sim \biggl( {k_B\over{m_p^4}}\, 
{\tau_{\rm tot}\over{\sigma_{SB}}} \, 
\beta_{\rm dyn}^{-1} \, \Omega^{-5} \, \Sigma \biggr)^{1/6} \, , 
\eeq
\beq
\rho_{0} \sim \Sigma/z_0    \sim \biggl( {k_B\over{m_p^4}}\, 
{\tau_{\rm tot}\over{\sigma_{SB}}} \, 
\beta_{\rm dyn}^{-1} \, \Omega^{-5} \, \Sigma^{-5} \biggr)^{-1/6} \, , 
\eeq

We can then also use equation~(\ref{eq-Mdot}) to express the mass accretion rate in terms of $\Sigma$ and~$\Omega$: 
\begin{eqnarray}
\dot{M} &\sim & \biggl({k_B\over{m_p}}\biggl)^{4/3} \, 
\biggl({\tau_{\rm tot}\over{\sigma_{SB}}}\biggr)^{1/3} \, 
\beta_{\rm dyn}^{-4/3} \, \Omega^{-2/3}\, \Sigma^{4/3}  \\
&\sim &
\biggl({k_B\over{m_p}}\biggl)^{4/3} \, 
\biggl({\kappa\over{\sigma_{SB}}}\biggr)^{1/3} \, 
\beta_{\rm dyn}^{-4/3} \, \Omega^{-2/3}\, \Sigma^{5/3}  \, .
\end{eqnarray}

As one can easily check these scalings are exactly the same as those obtained by \cite{Shakura_Sunyaev-1973} [see their eqs.~(2.16)] in the appropriate limit (gas-pressure dominated disk, constant opacity, etc.). 

Thus, we believe that much of the analysis developed in the main part of this paper can also be applied to the zero-net-flux case, resulting in an essentially very similar vertical disk structure (represented by the temperature and density profiles) and enabling one to recover the classical \cite{Shakura_Sunyaev-1973} scalings for the key disk parameters. While in this paper we do not want to make any strong claims regarding the coronal fraction $f$ of the accretion power in the zero-net-flux case, we nevertheless note that a typical value of $\beta_0 \sim 100$ expected in this case implies, in conjunction with equation~(\ref{eq-f-beta_0}), a relatively high (tens of percent, which is consistent with AGN observations) and universal (independent of any system parameters) value of~$f$.


\vskip 30 pt

\section{Conclusions}
\label{sec-conclusions}

In this paper we considered the problem of the vertical thermal structure of a thin gas-pressure-dominated accretion disk heated by the dissipation of MRI turbulence and cooled by optically thick radiative cooling. We also considered the question of the fraction of the overall accretion energy that is transported by buoyantly rising magnetic loops into the tenuous corona lying above the disk. 

In the main part of the paper, we focused on the non-zero-net vertical flux case, in which the disk is threaded by an externally imposed vertical magnetic field, $B_0$. Because we are interested in an MRI-active disk, we considered the case where this field is relatively weak, 
$B_0^2/8\pi \ll P_0$, where $P_0$ is the gas pressure in the middle of the disk. At the same time, we assumed that the mean vertical field $B_0$ is stronger than the dynamo-generated magnetic field $B_{\rm dyn}$, which, according to numerical simulations, is expected to be of order $B_{\rm dyn}^2/8\pi \sim 10^{-2}\, P_0$. We argued that, under these assumptions, the volumetric heating rate, $Q$, due to the dissipation of the MRI turbulence (which we tentatively associate with the disruption of MRI channel modes by parasitic instabilities, \cite{GX-1994}) should scale as $\Omega B_0^2/8\pi$; in particular, it should be independent of the height $z$ inside most of the disk. It is particularly important that the $z$-profile of $Q$ does not trace the profiles of the gas density or pressure. Making use of this finding and assuming, in addition, that the opacity $\kappa$ is also constant in~$z$, we then were able to solve analytically the combined set of equations governing the vertical structure of the disk --- the hydrostatic pressure balance, the energy conservation, and the optically thick radiative transfer equation. As a result, we were able to obtain the $z$-profiles of the the gas temperature and density inside the disk: 
$T(z) = T_0\, (1-z^2/z_0^2)$, $\rho(z) = \rho_0 \, (1-z^2/z_0^2)^3$, where $T_0$ and $\rho_0$ are the values at the disk midplane, $z=0$, and $z_0$ is the effective thickness of the disk 
[see eqs.~(\ref{eq-T-z}) and~(\ref{eq-rho-z})]. 

We were also able to to determine all the key disk parameters, such as the midplane temperature, density, and pressure, and the disk thickness~$z_0$, in terms of the governing input parameters in this problem: the external vertical magnetic field~$B_0$, the surface density $\Sigma$, the disk rotation rate~$\Omega$, and the opacity$~\kappa$. This enabled us to evaluate the scaling of the mass accretion rate $\dot{M}$ with these parameters (see \S~\ref{sec-Mdot}) and formulate the equation governing the time evolution of the radial distribution of mass across the disk, $\Sigma(r,t)$. 

We then also examined how and where various assumptions on which our disk model is based --- $\beta(z) \gg 1$, $l_{\rm ph} \ll H$, etc. --- break down as one approaches the disk's surface. This enabled us to address the question of the coronal power fraction. In our view, the coronal power is governed by the competition between various parasitic instabilities disrupting the primary MRI channel modes: the GX94 instabilities leading to fully developed local MHD turbulence whose dissipation heats the disk locally, and the Parker instability that pumps the Poynting flux into the ADC by buoyantly rising magnetic loops. The magnetic energy of these loops is then dissipated by reconnection in the corona.  In our model, the resulting corona fraction turns out to be relatively insensitive to most input parameters, scaling as $f\sim \beta_0^{-1/5}$. The practical consequence of this conclusion is that it is difficult to avoid a sizable (tens of percent) fraction of the accretion power to be released in the overlying corona. 
The part of this dissipated energy that goes to the electrons powers the observed coronal emission, whereas a significant part of the ion coronal energy may actually be transported back to the disk by ions streaming along the closed field lines and get deposited in the dense disk through ion-ion collisions. Finally, some of the energy dissipated by reconnection involving open field lines may power outflows (winds and jets) along these open field lines. 

Finally, in \S~\ref{sec-zero-net-flux}, we turned our attention to the case of a disk with a zero net flux. This is the case where the next externally imposed vertical field~$B_0$ is either absent altogether or small compared to the MRI dynamo-generated field~$B_{\rm dyn}$. In this case, the entire steady-state vertical structure of a thin gravitationally-stratified disk should be determined solely by $\Sigma$, $\Omega$, and $\kappa$ since there are no other parameters in the problem (ignoring external irradiation).  As numerical simulations show, in this case one expects an effective large-scale MRI dynamo to produce a sizable large-scale (comparable or larger than the disk pressure scale-height~$H$) magnetic field $B_{\rm dyn}^2/8\pi \sim \beta_{\rm dyn}^{-1}\, P_0 $ with $\beta_{\rm dyn} \simeq 10^2$. We argued that since the magnetic energy density is roughly uniform in~$z$ inside the disk (perhaps up to the equipartition height at which $P(z) = B_{\rm dyn}^2/8\pi\simeq 10^{-2}\, P_0$), then the volumetric energy dissipation rate of the MHD turbulence should also be roughly uniform, just as it is for the finite net flux case. This picture is in fact supported by numerical simulations of stratified disk \cite[e.g.,][]{Miller_Stone-2000, Hirose_etal-2006}. We then argued that we can apply the analysis developed in \S~\ref{subsec-structure} to the zero-net-flux case. As a result, we recover the same vertical profiles for the gas temperature and density, i.e., $T(z) =T_0\, (1-z^2/z_0^2)$, and $\rho(z) = \rho_0\, (1-z^2/z_0^2)^3 $ [see equations~(\ref{eq-T-z}) and~(\ref{eq-rho-z})]. These profiles seem to be a good agreement with those obtained in full numerical simulations by~\cite{Hirose_etal-2006}. Furthermore, we find our scalings of the main accretion disk parameters (the midplane values of the temperature and density, the disk thickness~$z_0$, and the mass accretion rate~$\dot{M}$, etc.) to be the same as those in the classic paper by \cite{Shakura_Sunyaev-1973} for the regime under consideration (constant opacity, gas pressure-dominated disk). 

While the theoretical model presented in this paper is successful in being able to provide a set of concrete, physically motivated predictions, it still has to be viewed just as an idealized conceptual toy model. It relies on a number of simplifying assumptions --- e.g., gas-pressure domination, a single constant opacity, external irradiation --- that preclude its direct application to real astrophysical systems. For example, the inner regions of optically-thick BH accretion disks are dominated by the radiation pressure, which is completely ignored in the present study. In addition, the magnetic field generated by the MRI dynamo is either completely ignored (in the first half of the paper, where we consider a disk threaded by an external magnetic field) or is treated in a simplified fashion (in \S~\ref{sec-zero-net-flux}). Perhaps for these reasons the present model is not able to address the questions of spectral state transitions in galactic BH binaries and, in particular, to explain the relatively low level of X-ray coronal activity in the high-soft state. Future theoretical studies should develop generalizations of the present work to take into account radiation pressure, more realistic opacities, and, perhaps, external irradiation. One can also envision a more rigorous and, perhaps, more accurate model of MRI saturation and MHD turbulent dissipation; the development of such a model should benefit from detailed direct comparisons with numerical simulations.

In conclusion, we believe that the future of accretion disk studies lies in incorporation of more realistic radiation and thermal physics --- i.e., more accurate treatment of the disk thermodynamics and radiative cooling processes. We hope that the present paper will help stimulate and pave the way for such studies using advanced numerical simulations, along the lines of the studies by \cite{Hirose_etal-2006, Hirose_etal-2009, Shi_etal-2010, Blaes_etal-2011}. 
Eventually, we hope, this line of research will reach the state of maturity at which it is able to provide meaningful predictions for observations and explain important observational facts.


\begin{acknowledgments}

I am very grateful to Prof. Jeremy Goodman for many fruitful and stimulating discussions. I also would like to thank M. Begelman, J. Goodman, and J. Simon for useful comments on this manuscript. 
This work is supported by National Science Foundation grant 
No.\, PHY-0821899 (PFC: Center for Magnetic Self-Organization 
in Laboratory and Astrophysical Plasmas) and by NASA under grant 
No.~NNX11AE12G.

\end{acknowledgments}





\end{document}